# Classification of Titanic Passenger Data and Chances of Surviving the Disaster

John Sherlock, Manoj Muniswamaiah, Lauren Clarke, Shawn Cicoria
Seidenberg School of Computer Science and Information Systems
Pace University
White Plains, NY, US
{js20454w, mm42526w, lc18948w}@pace.edu, shawn@cicoria.com

*Abstract –* While the Titanic disaster occurred just over 100 years ago, it still attracts researchers looking for understanding as to why some passengers survived while others perished. With the use of a modern data mining tools (Weka) and an available dataset we take a look at what factors or classifications of passengers have a persuasive relationship towards survival for passengers that took that fateful trip on April 10, 1912. The analysis looks to identify characteristics of passengers - cabin class, age, and point of departure – and that relationship to the chance of survival for the disaster.

## I. INTRODUCTION

The Titanic was a ship disaster that on its maiden voyage sunk in the northern Atlantic on April 15, 1912, killing 1502 out of 2224 passengers and crew[2]. While there exists conclusions regarding the cause of the sinking, the analysis of the data on what impacted the survival of passengers continues to this date[2,3]. The approach taken is utilize a publically available data set from a web site known as Kaggle[4] and the Weka[5] data mining tool. We focused on decision tree based and cluster analysis after data review and normalization.

### A. Kaggle – Predictive Modeling and Analytics

Kaggle offers businesses and other entities crowd-sourcing of data mining, machine learning, and analysis. Sometimes offering prizes (for example there had been a $200,000 prize being offered from GE through Kaggle in a competition[1]).

### B. Weka - Waikato Environment for Knowledge Analysis

The Weka tool provides a collection of machine learning and data mining tools. Freely available built upon Java which allows it to run on platforms that support Java. It's maintained and supported primarily by researchers at the University of Waikato.

## II. DATA AND METHODOLOGY

### A. Sample Data from Kaggle

The following is a representation of the test dataset provided in a comma separated value (CSV) format from Kaggle and 891 rows of data (a subset of the entire passenger manifest). The file structure with example rows is listed in the following 3 tables.

Table I. Kaggle Titanic Sample Data

| passengerid | survived | pclass | name | sex |
|---|---|---|---|---|
| 1 | 0 | 3 | Braund, Mr | male |
| 2 | 1 | 1 | Cummingss, Mrs | female |
| 3 | 1 | 3 | Heikkinen, M | female |

Table II. Kaggle Titanic Sample Data (Continued)

| age | sibsp | parch | ticket |
|---|---|---|---|
| 22 | 1 | 0 | A/5 2117 |
| 38 | 1 | 0 | PC 17599 |
| 26 | 0 | 0 | STON/O2 |

Table III. Kaggle Titanic Sample Data (Continued)

| fare | cabin | embarked |
|---|---|---|
| 7.25 |  | S |
| 71.2833 | C85 | C |
| 7.925 |  | S |

### B. Data Normalization

The dataset was modified to create nominal columns from some of the numeric columns in order to facilitate usage in Weka for Tree analysis and simple cluster analysis.

The modification is done to facilitate usage in Weka for tree analysis and simple cluster analysis. The following table identifies the conversions and other modifications.

TABLE I. KAGGLE DATASET NORMALIZED DATA TYPES

| Field | Modification | Comment |
|---|---|---|
| PassengerID | Ignored | Not needed |
| Survived | Converted to NO/YES | Needed nominal identifier |
| Pclass | Removed -> created class column instead | Needed nonminal identifier |
| Class | New Column | Simple calculation based upon 'pclass' |
| Agegroup | Formula based; some values not supplied. But ended up with 4 groups other than Unknown (Child, Adolescent, Adult, Old) | Arbitrarily did the following:<br>=IF(F2="", "Unk",IF(F2<10, "Child", IF(F2<20, "Adolescent", IF(F2<50, "Adult", "Old")))) |
| Ecode | Removed -> created class Embarked | Needed nominal identifier |
| Embarked | New column that converted Ecode to the real name of the departure point for the passenger | |

*C. Normalized Analysis Dataset*

Upon conversion, the final dataset utilized for the analysis in the Weka tool is illustrated below with the first few rows shown.

TABLE II. NORMALIZED DATASET EXAMPLE

| PassengerId | Survived | Pclass | Class |
|---|---|---|---|
| 1 | No | 3 | 3rd |
| 2 | Yes | 1 | 1st |
| 3 | Yes | 3 | Erd |

TABLE III. NORMALIZED DATASET EXAMPLE (CONTINUED)

| Sex | Age | AgeGroup | Ecode | Embarked |
|---|---|---|---|---|
| male | 22 | adult | S | Southhamptom |
| Female | 38 | adult | C | Cherbourg |
| Female | 26 | adult | S | Southhamtom |

*D. Weka ARFF file Format*

The table is then converted and saved into the Weka Attribute-Relation File Format (ARFF). The ARFF file used is represented in appendix E. The key characteristics of the ARFF file format in order to facilitate the data exploration in the Weka tool is the identification of the data types and within those fields the order of the nominal values.

### III. DATA ANALYSIS

*A. Decision Tree Classification*

Using Weka, we generated a J48[6] Tree (C.45 implementation) which resulted in the classifier output represented in appendix G - J48 Classifier Output. The J48 Tree diagram shown in figure 2 below illustrates the classification path that the data suggests.

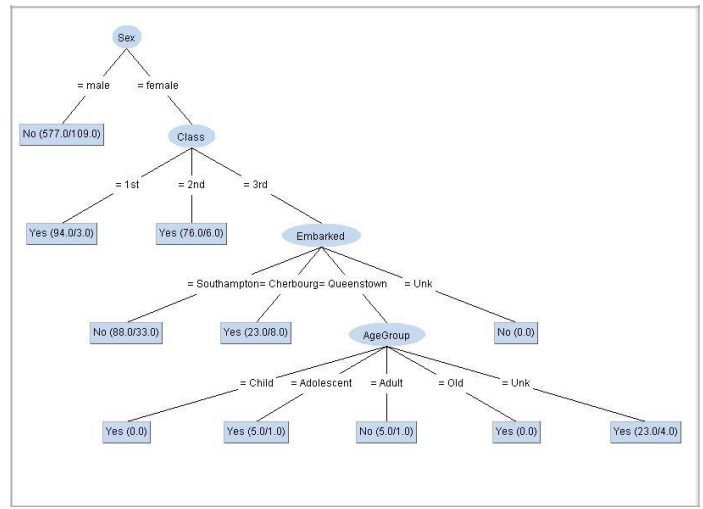

Fig. 1. J48 Classifer Diagram

*B. J48 Classifier - Initial Conclusions*

Based upon the outcome of the J48 analysis it was clear that the most significant association with regards to survival was related to Sex; in that just being Female was the most significant classifier. We then reviewed the cluster analysis for further relationships.

*C. Simple K Means Cluster Analysis*

Clustering the data based upon classifications and use of clustering analysis simple associations may be understood from the data. While an association might be strong through this analysis, the true cause and effect cannot be concluded.

*D. Simple K Means Output*

For our cluster analysis, we chose the Simple K Means, just for simplicity. The Simple K Means text Output is included in appendix H. The visualizations are also shown in the following sections.

Using the cluster diagram we can visually analyze the clusters for relationships within the dataset. The strength of the classification and clustering is shown visually as well as within the text output. This clustering relationship may be used to conclude that some relationship exists, but not cause-and-effect.

*E. Survived vs. Sex*

Quite dramatically visually we see that sex of the passenger shows significant clustering around survival chances. This had been also shown in the J48 tree. Figure 2 below illustrates the significant clustering of Sex vs. chance of survival. Whether this is anticipated or not is something that would require further corollary analysis within social sciences as to why one Sex may fare better in these traumatic situations.

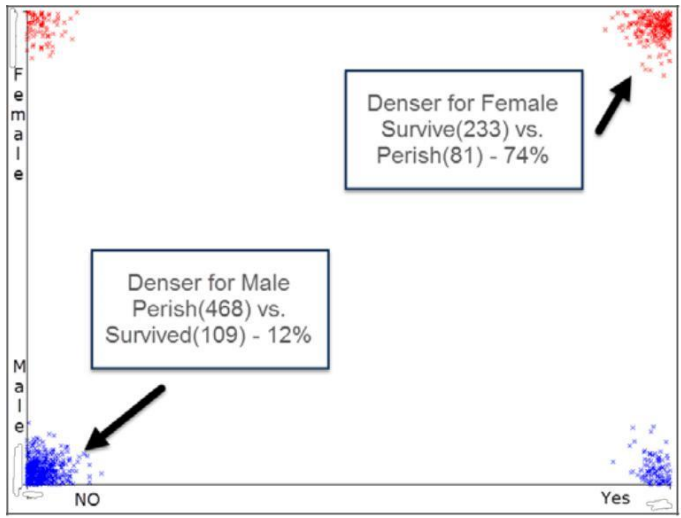

Fig. 2. Simple K Means – Survived vs. Sex Classification

*F. Survived vs. Class*

Perhaps not surprisingly, cabin class had significant clustering with the lower tiered cabins showing significant weight towards non-survival. This is shown in figure 3 below with a fairly dominant clustering for those in $3^{rd}$ class that did not survive. And somewhat clear clustering for those in $1^{st}$ class surviving. We can make conjectures about this result, perhaps the physical location or other facts about how passengers were able to freely or not freely move about the ship. However, we cannot drawl strong conclusions or inferences from this data alone.

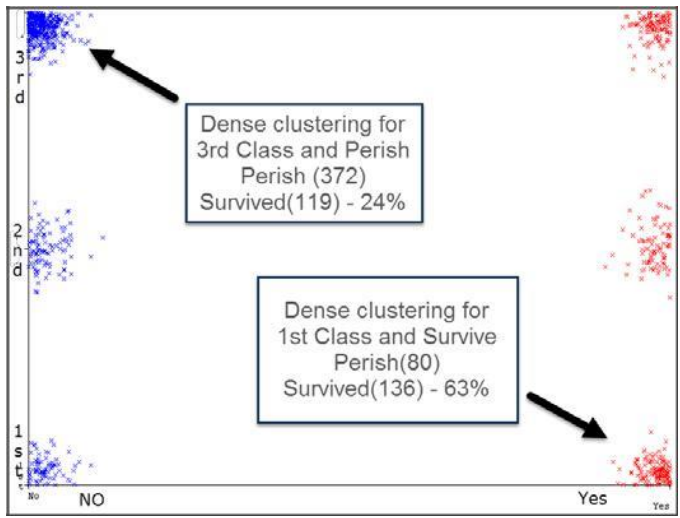

Fig. 3. Simple K Means Visualization – Survived vs. Class

*G. Survived vs. Age Group*

Our data normalization arbitrarily bucketed passenger data into various nominal groups. Amongst these groups, again not surprisingly the adults – age 20 – 49 – were amongst those that perished. Figure 4 below doesn't have as great of a visual clustering as the prior two – sex or age. Our approach to age buckets was a generalization. Further analysis could be made with more granular or genealogically defined age groups to draw any further potential relationships that might exist.

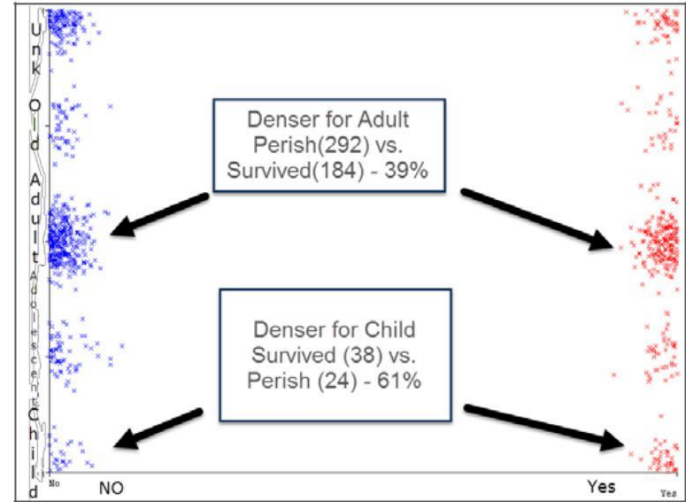

Fig. 4. Simple K Means Visualization – Survived vs. Age Group

*H. Survived vs. Embarked*

Finally, the analysis identified that point of embarking of the passengers was also an indicator of survival rate, although not as strong. What has not been done is the association of point of embarking with cabin class – e.g. did $3^{rd}$ class passengers mostly embark at Southampton. This is shown in figure 5 below.

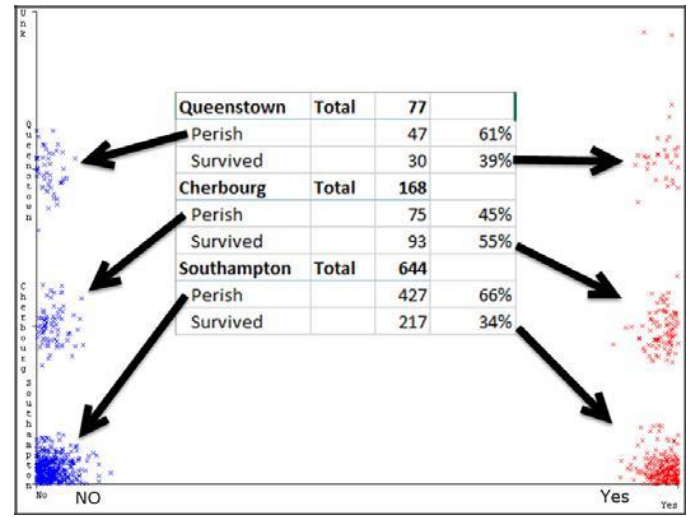

Fig. 5. Simple K Means Visualization – Survived vs Embarked

Finally, indication on the cluster diagram what the class of the passengers is then shows a strong association between class and point of departure. Thus, these 2 features seem related in some manner and probably not independent.

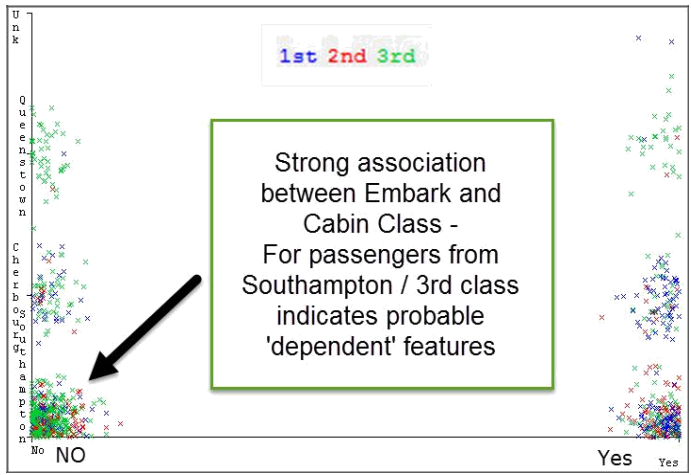

Fig. 6. Simple K Means – Survived vs. Embarked – Class coloring

## IV. FINAL OBSERVATIONS

Sex clearly had the most significant relationship demonstrated within the dataset for survival rate. In addition, the J48 classifier, using the test data set resulted in ~ 81% correctly classified instances. In comparison to the Kaggle competition as of the writing of this paper, this put the model in about 43rd place.

## V. SUMMARY

The Weka tool, while powerful, requires coaxing of the data into a more amiable format to facilitate tool usage and classification approaches work. This was a good learning experience for the team. At first, we had chosen baseball statistics but quickly became overwhelmed with the number of attributes and the size of the data sets. Since conversion of the data from numeric types to classifiers was an onerous task, the baseball statistics were tossed out and we searched for another dataset. Weka and the algorithms required nominal values for classifiers instead of numeric values.

We discovered the dataset from Kaggle and with simple manipulation we were able to arrive at a quite compatible dataset in ARFF format (Weka native format) that worked well and provided quite significant results that demonstrated which classes of passengers had the greatest impact on survivability.

## VI. FUTURE RESEARCH

The dataset utilized represented a subset or "test" dataset used for the Kaggle competition. With the complete dataset the model can be validated and some of the same conclusions or relationships verified. Additionally, looking at some of the other cross classification dependencies – such as cabin class and embark location – to eliminate unnecessary classifiers should be done.

## VII. REFERENCES

[1] GE, "Flight Quest Challenge," Kaggle.com. [Online]. Available: https://www.gequest.com/c/flight2-main. [Accessed: 13-Dec-2013].

[2] "Titanic: Machine Learning from Disaster," Kaggle.com. [Online]. Available: https://www.kaggle.com/c/titanic-gettingStarted. [Accessed: 13-Dec-2013].

[3] Wiki, "Titanic." [Online]. Available: http://en.wikipedia.org/wiki/Titanic. [Accessed: 13-Dec-2013].

[4] Kaggle, Data Science Community, [Online]. Available: http://www.kaggle.com/ [Accessed: 13-Dec-2013]

[5] Weka 3: Data Mining Software in Java, [Online]. Available: http://www.cs.waikato.ac.nz/ml/weka/ [Accessed: 13-Dec-2013]

[6] C4.5 Algorithm, Wikipedia, Wikimedia Foundation, [Online]. Available: http://en.wikipedia.org/wiki/C4.5_algorithm, [Accessed: 13-Dec-2013]

## VIII. APPENDICES

*A. Sample Data From Kaggle – Initial Dataset*

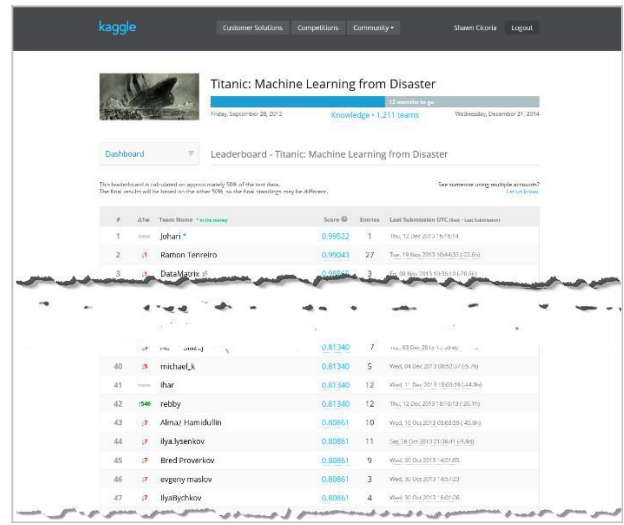

*B. Normalized Dataset based upon Kaggle Data*

*C. Kaggle Competition – Titanic Disaster Leaderboard*

*D. Weka Screens – Main Preprocess View*

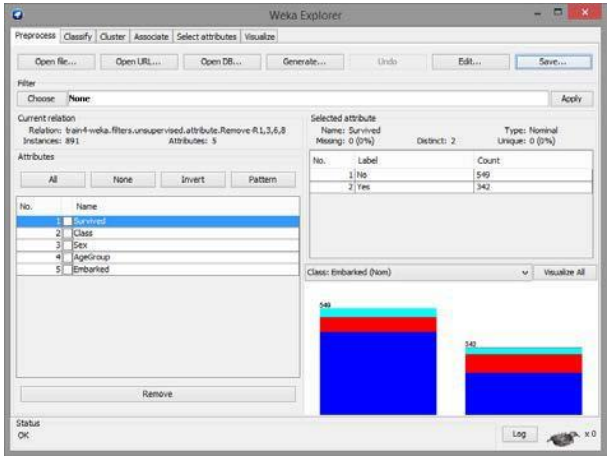

*E. Weka Screens – J48 Classify View*

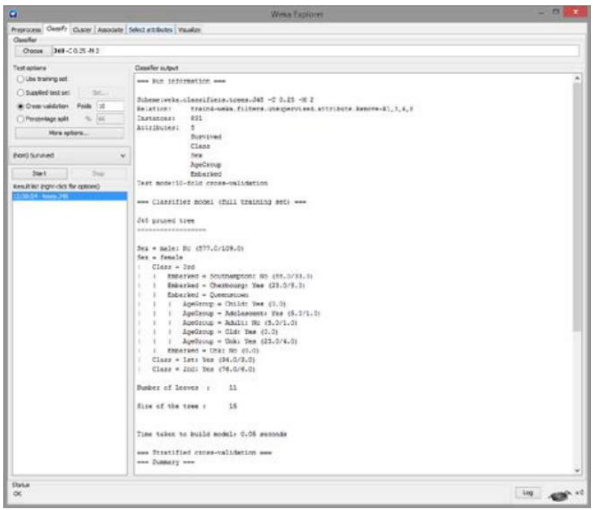

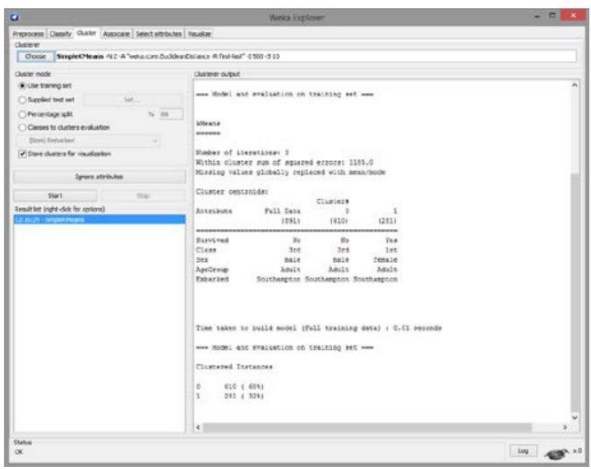

*F. Normalized Dataset in ARFF format*

> @relation 'train4-weka.filters.unsupervised.attribute.Remove-R1,3,6,8'
>
> @attribute Survived {No,Yes}
> @attribute Class {1st,2nd,3rd}
> @attribute Sex {male,female}
> @attribute AgeGroup {Child,Adolescent,Adult,Old,Unk}
> @attribute Embarked {Southampton,Cherbourg,Queenstown,Unk}
>
> @data
> No,3rd,male,Adult,Southampton
> Yes,1st,female,Adult,Cherbourg

*G. J48 Classifier Output (3 parts)*

```
=== Run information ===
Scheme:weka.classifiers.trees.J48 -C 0.25 -M 2 Relation:
   train4-
weka.filters.unsupervised.attribute.Remove-
R1,3,6,8
Instances:891
Attributes: 5 Survived
           Class
           Sex
           AgeGroup
           Embarked

Test mode:10-fold cross-validation
```

```
=== Classifier model (full training set) ===
J48 pruned tree
------------------
Sex = male: No (577.0/109.0)
Sex = female
|   Class = 3rd
|   |   Embarked = Southampton: No (88.0/33.0)
|   |   Embarked = Cherbourg: Yes (23.0/8.0)
|   |   Embarked = Queenstown
|   |   |   AgeGroup = Child: Yes (0.0)
|   |   |   AgeGroup = Adolescent: Yes (5.0/1.0)
|   |   |   AgeGroup = Adult: No (5.0/1.0)
|   |   |   AgeGroup = Old: Yes (0.0)
|   |   |   AgeGroup = Unk: Yes (23.0/4.0)
|   |   Embarked = Unk: No (0.0)
|   Class = 1st: Yes (94.0/3.0)
|   Class = 2nd: Yes (76.0/6.0)

Number of Leaves  :     11
Size of the tree :     15
Time taken to build model: 0.05 seconds

=== Stratified cross-validation ===
```

```
=== Stratified cross-validation ===
=== Summary ===
Correctly Classified Instances         722
81.0325 %
Incorrectly Classified Instances       169
18.9675 %
Kappa statistic                          0.5714
Mean absolute error                      0.2911
Root mean squared error                  0.385
Relative absolute error                 61.5359 %
Root relative squared error             79.1696 %
Total Number of Instances              891
=== Detailed Accuracy By Class ===
         TP Rate    FP Rate   Precision   Recall
         0.953      0.418     0.785       0.953
         0.582      0.047     0.884       0.582
W.Avg.   0.81       0.276     0.823       0.81
F-Measure  ROC Area   Class
0.861      0.783      No
0.702      0.783      Yes
0.8        0.783

=== Confusion Matrix ===
   a    b   <-- classified as
 523   26 |   a = No
 143  199 |   b = Yes
```

## H. Simple K Means Clustering Output

```
=== Run information ===
Scheme:weka.clusterers.SimpleKMeans
   -N 2 -A "weka.core.EuclideanDistance -
   R first-last" -I 500 -S 10
Relation:train4-weka.filters
   .unsupervised.attribute
      .Remove-R1,3,6,8
Instances:891
Attributes: 5 Survived
            Class
            Sex
            AgeGroup
            Embarked

Test mode:evaluate on training data

=== Model and evaluation on training set ===
kMeans
======
Number of iterations: 3
Within cluster sum of squared errors: 1185.0
Missing values globally replaced with mean/mode
Cluster centroids:
                             Cluster#
 Attribute    Full Data            0            1
               (891)            (610)        (281)
==========================================
Survived           No              No          Yes
Class             3rd             3rd          1st
Sex              male            male       female
AgeGroup        Adult           Adult        Adult
Embarked   Southampton Southampton Southampton
Time taken to build model
(full training data) : 0.01 seconds
=== Model and evaluation on training set ===
Clustered Instances
0      610 ( 68%)
1      281 ( 32%)
```